\documentstyle[12pt]{article}
\begin{document}

\newcommand{\parzialet}[1]   {  \frac{ \partial{#1} }{\partial{t}} }
\newcommand{\derivatax}[1] {  \frac{        d{#1} }{        dx} }

\newcommand{\traccia}{ {\rm Tr} }
\newcommand{\reale}  { {\rm Re} }

\newcommand{\ca}{ {\cal A} }
\newcommand{\ch}{ {\cal H} }
\newcommand{\cm}{ {\cal M} }

\newcommand{\be}{\begin{equation}}
\newcommand{\ee}{\end{equation}}

\title{The String Tension in Gauge Theories: a Suggestion for a
       New Measurement Method}

\author{
Enzo MARINARI$^{(1,2)}$, Maria Luigia PACIELLO$^{(3)}$,\\
Giorgio PARISI$^{(1,4)}$ and Bruno TAGLIENTI$^{(3)}$\\[0.3em]
{\normalsize $^{(1)}$: Dipartimento di Fisica and Infn,}\\
{\normalsize Universit\`a di Roma {\em Tor Vergata},}\\
{\normalsize Viale della Ricerca Scientifica, 00133 Roma (Italy)}\\
{\normalsize $^{(2)}$: Physics Department and NPAC,}\\
{\normalsize Syracuse University, Syracuse, NY 13244 (USA)}\\
{\normalsize $^{(3)}$: Infn Sezione di Roma}\\
{\normalsize Universit\`a di Roma {\em La Sapienza},}\\
{\normalsize Piazzale Aldo Moro, 2 - 00185 Roma (Italy)}\\
{\footnotesize
  marinari, paciello, parisi, taglienti @roma1.infn.it}}

\maketitle
\vfill
{\footnotesize $^{(4)}$: Address after November 1st:
 Dipartimento di Fisica and Infn -
 Universit\`a di Roma {\em La Sapienza},
 Piazzale Aldo Moro, 2 - 00185 Roma (Italy)}

\begin{flushright}
  { ROMA 92 - 903}\\
  {\bf hep-lat/9210021 }
\end{flushright}
\newpage
\begin{abstract}

We discuss a new method for testing confinement and  measuring the string
tension (in the Coulomb gauge). Our numerical simulations demonstrate that the
problems related to Gribov copies are not harmful and that the method is
effective in the case of pure gauge Q.C.D.. We discuss the relevance of the
method for studying gauge theories coupled to fermionic matter.

\end{abstract}

\vfill

\newpage

In this note we introduce a new method (following an {\em en passant} remark
of ref. \cite{GIORGIO}) for measuring the string tension $\sigma^2$ in gauge
theories, and to establish a criterion for confinement. We show numerically
the validity of this approach, and we discuss its relevance toward the
simulation of fully coupled Q.C.D..

Our approach will be based on the use of Coulomb gauge; we will deal with the
gauge fixed lattice theory, in a gauge that is smooth at fixed euclidean
time, i.e. where the spatial gauge fields are brought, by gauge
transformations, as close as possible to the identity.

We will consider in the following an $SU(3)$ gauge theory defined on a lattice
of volume $L^3$ and time extent $T$, with periodic boundary conditions in the
$4$ dimensions.

The {\em gaugeon} of length $n$ is defined by

\be
  G_n(\vec{x},t_0) \equiv \prod_{t=t_0}^{t_0+n}
  \big [ U_t(\vec{x},t) \big ] \ ,
\ee

and we integrate over a spatial $2$-plane (going to zero $2$-momentum) by
setting

\be
  G_n(x_\alpha,t_0) \equiv \frac{1}{L^2}
  \sum_{\vec{x}_{\beta},\vec{x}_{\gamma}}
  G_n(\vec{x},t_0)\ ,
\ee
where $\alpha$, $\beta$, $\gamma$ $=$ $1$, $2$, $3$, $\alpha$ $\neq$ $\beta$
$\neq$ $\gamma$. We compute the zero $2$-momentum correlation functions by
defining

\be
  \protect\label{E_CN}
  C_n(d) \equiv \frac{1}{3LT}
  \sum_{\delta=1,2,3} \sum_{t_0=1,T} \sum_{x_\alpha=1,L}
  \langle G_n(x_\delta,t_0) G_n^\dagger(x_\delta+d,t_0) \rangle \ .
\ee

Let us introduce the point by elaborating, in the form discussed in ref.
\cite{GIORGIO},
a point originally developed by Ferrari and Picasso$^{\cite{FERPIC}}$. The
argument hints the relevance of measuring, in the Coulomb gauge, correlation
functions of time-like gauge fields at the same time (and different spatial
points), of the type (\ref{E_CN}).

The point of view suggests that {\em the photon can be seen as the Goldstone
boson of the gauge symmetry}, and that the instantaneous potential
$\frac{1}{x}$ (the Coulomb potential in Coulomb gauge) can bee seen as
originated from the exchange of a Goldstone boson. For understanding the
point let us consider our lattice theory, which has an invariance
$SU(3)^{L^3T}$. This is the gauge symmetry of the lattice theory, and the
gauge group is somehow too large to be broken from Goldstone bosons.

We can anyhow gauge fix our theory, and reduce the symmetry. Let us fix
Coulomb gauge, by maximizing the expression

\be
  \protect\label{E_GAUGE}
  \sum_i \sum_{\mu=1}^3 \reale \{ \traccia [ U_\mu(x) ] \} \ .
\ee

We can get rid of Gribov copies in many different ways.  We can define the
expectation values by averaging over  all copies with equal probability, or we
can  choose, with lot of work, the global maximum (which is generically
unique).
We could also assign to each copy a different weight which is proportional to
its basin of attraction in the algorithm we are using to fix the gauge. We note
that for similar gauge fixing algorithms the basins of attraction are quite
similar.

Independently from the method we use to deal with Gribov copies,
the crucial point is that now there is a residual symmetry. Gauge
transformations which only depend on time but are space independent leave the
quantity (\ref{E_GAUGE}) invariant. This symmetry is a global $SU(3)$ for
each time slice, i.e. the total residual symmetry is $SU(3)^T$. But on a
fixed time slice (a sensible entity to consider in Coulomb gauge) now we
have a global symmetry, which in the $V \to \infty$ can by broken generating a
Goldstone boson. In the $V \to \infty$ limit the symmetry will be indeed
broken in the Coulomb phase (where we will have a Goldstone boson for each
time slice, and the expected propagator), while it will be preserved in the
confined phase. $C_1(d)$ will tend to a constant for $d \to \infty$ in the
Coulomb case, while it will decay exponentially in the confined phase.

This physical picture leads us to suggest to use $C_n(d)$ in order to measure
the string tension. We expect that for $n$ and $d$ large
enough $C_n(d)$ will decay, in the confined phase, with a behaviour

\be
  e^{-\sigma^2 n d}\ .
\ee

There are two ways which are usually employed to measure the string tension
$\sigma^2$ and to distinguish between the confined and the deconfined
phase.  One is based on the measurement of large Wilson loops (the
original {\em Creutz ratios}), while the other is based on measuring
correlation functions of Polyakov loops. Statistical improvements, like for
example the use of {\em smeared} looppy observables, turn out to be crucial
(related to the fact we are working in a critical limit, where a correlation
length is diverging).

In a pure gauge theory the expectation value of a Wilson loop of area
$ A=B\times H$ behaves as $e^{-A}$ if the theory is confined, and as $e^{-L}$
if the theory is deconfined. If we couple the theory to fermions we can close
a fermion loop only paying a price proportional to the loop length, and we
get again an $e^{-L}$ decay. So the Wilson loop ceases to be a good indicator
when we deal with the full theory.

The Polyakov loop correlation function at distance $d$ behaves for large $d$
as $e^{-\sigma^2 L d}$ in a confined pure gauge theory, and gets a
non-zero connected part when the pure theory deconfines. Also in this case
the fully coupled theory does not acknowledge a difference between the two
phases, since also in the confined phase the fermion loops give a non zero
expectation value to the loop-loop correlation function. The two most popular
ways used to determine the string tension $\sigma^2$ and to
distinguish between the two phases are not effective in the
case of the theory coupled to fermions.

On the contrary we expect the {\em gaugeon-gaugeon} correlation functions
$C_n(d)$ behave asymptotically (for large $d$) as  $e^{-f(n) d}$ both in the
pure gauge theory and in the theory coupled to fermions in the confined phase.
In the case of the pure gauge theory we expect $f(n)$ to coincide (for large
$n$) with $\sigma^2 n$. Here indeed the $U$ cannot take an expectation  value
if
the symmetry is unbroken. The method can be used both in the pure gauge and in
the fermionic theory, and is likely to be a very effective method in both
cases.
In the following we will discuss a pure gauge numerical simulation in which we
demonstrate its effectiveness.

We have analyzed $100$ configurations on a $10^3 \times 20$ lattice and $412$
configurations on a $10^3 \times 6$ lattice, both at $\beta=5.8$. They have
been separated (after $2000$ thermalization sweeps) of $500$ sweeps of an $8$
hit Metropolis updating scheme. Coulomb gauge has been fixed by using an
over-relaxed procedure. On each independent gauge configuration we have gauge
fixed ten times, starting from $10$ different randomly gauge transformed
samples. We were interested to check if Gribov copies can have an influence of
such a quantity (since it is computed in a gauge fixed environment). So we
have averaged separately the configurations which turned out to have a maximum
value of (\ref{E_GAUGE}), the medium ones and the minimum ones. We have
independently computed the rate of the $3$ decays, and in the limit of our
statistical error we have not seen any difference between them.

In fig. 1 we show the effective mass estimator, as defined from the logarithm
of the ratio of two $C_n(d)$  (with the corrections needed from the presence
of periodic boundary conditions) for our largest lattice. We have points
for the ration of distance $1$ and $2$, $2$ and $3$ and $3$ and $4$. The lines
give our best fit, of the form

\be
  \sigma^2_{(n)} = \sigma^2_{(\infty)} + \frac{c_1}{n}\ ,
\ee
which turns out to be perfect in all cases. Here measuring directly an
estimator for the string tension (also at distance $1$ and $2$, which is
however highly biased, since we are used local, non-smeared Wilson loops) is
impossible, since the lattice is too large (the time asymptotic result is of
order $0.1$, see for example \cite{ENZO}). In order to stress the very good
linearity of our data as a function of the {\em gaugeon} size $n$ we plot
the effective mass as a function of $n$ in Fig. 2.

In Figs. 3 and 4 we plot the same data for the $10^3 \time 6$ lattice. Here
we can compute the true estimator for $\sigma^2_{effective}$ at distance $1$
over $2$, and we find that there is a small systematic difference from the
curve extrapolated by using the non-gauge invariant {\em gaugeons}. We expect
such a small systematic effect, which will tend to zero in the continuum
limit, since this is a gauge invariant measurement. This additional
systematic error has to be kept under control, but does not seem to be a
dangerous effect, already at a quite low value of $\beta$.

This result is very satisfactory, as far as the $n$-dependence of the effective
mass estimator at fixed distance seems to be under very good control.  In the
limit of our statistical precision The presence of Gribov copies seems to be
irrelevant as far as our results do not depend on criterion we have used to
choose the copies. Obviously in a practical implementation of the method one
will
limit is search to a single gauge fixed copy for each independent starting
gauge
configuration (where, at $\beta=5.8$, the decorrelation time for local
observables will be surely smaller than the one we used here to be sure to get
rid of all non-local correlations).

As far as we have been able (by using non-smeared Wilson loops)
to measure the effective mass only up to distance $3$ over $4$
the extrapolation of the effective mass estimator at large distance is
problematic. We can only notice that the estimator from times $3$ over
$4$ (extrapolated at large $n$) is about the $50\%$ higher than the
asymptotic value of the string tension: this is the best upper bound
to the value of $\sigma^2$ we have been able to obtain and is a quite
reasonable result. We expect the use of smeared operators to be very
effective in decreasing the error.
Checking how effective the method is when
dealing with smeared operators seems to be the next important step.

\vfill
\newpage

\section{Figure Captions}
  \begin{itemize}

    \item[Fig. 1] Distance dependent effective mass estimator as a function
                  of the {\em gaugeon} size $n$. $10^3 \times 20$ lattice.
                  Continuous lines are the best fits.

    \item[Fig. 2] As in fig. $1$, but as a function of $n^{-1}$.

    \item[Fig. 3] As in fig. $1$, but $10^3 \times 6$ lattice.

    \item[Fig. 4] As in fig. $1$, but $10^3 \times 6$ lattice and as a
                  function of $n^{-1}$.

\end{itemize}

\vfill


\vfill
\newpage

\begin{thebibliography}{99}
  \bibitem{GIORGIO}
    G. Parisi,
    {\em A Short Introduction to Numerical Simulations of Lattice Gauge
         Theories,}
    in {\em Ph\'enom\`enes Critiques, Syst\`emes Al\'eatories, Theories de
            Jauge,} p. 87,
    edited by K. Osterwalder and R. Stora
    (North-Holland, Amsterdam, The Netherlands 1986).
  \bibitem{FERPIC}
    R. Ferrari and R. Picasso, Nucl. Phys. {\bf B31} (1971) 316.
  \bibitem{ENZO}
    E. Marinari,
    Nucl. Phys. {\bf B} (Proc. Suppl.) {\bf 9} (1989) 209.
 \end{thebibliography}
\end{document}